\documentclass{article}

 \usepackage[preprint]{neurips_2026}

\usepackage[utf8]{inputenc} % allow utf-8 input
\usepackage[T1]{fontenc}    % use 8-bit T1 fonts
\usepackage{hyperref}       % hyperlinks
\usepackage{url}            % simple URL typesetting
\usepackage{booktabs}       % professional-quality tables
\usepackage{amsfonts}       % blackboard math symbols
\usepackage{nicefrac}       % compact symbols for 1/2, etc.
\usepackage{microtype}      % microtypography
\usepackage[table]{xcolor}         % colors
\usepackage{natbib}
\usepackage{amsmath}
\usepackage{makecell}
\usepackage{svg}
\usepackage{wrapfig}
\usepackage{graphicx}
\usepackage{pifont}
\usepackage{multirow}
\usepackage{subcaption}
\newcommand{\xmark}{\ding{55}}
\title{QuantWeather: Quantile-Aware Probabilistic Forecasting for Subseasonal Precipitation}

\newcommand{\model}{QuantWeather}
\author{%
Lei Chen$^{1,2}$\thanks{These authors contributed equally to this work.},
Xinyu Su$^{1,2}$\footnotemark[1],
Xiaohui Zhong$^{1,2,3}$,
Hao Li$^{1,2,3}$\thanks{Corresponding author. \texttt{lihao\_lh@fudan.edu.cn}}\\
$^{1}$Artificial Intelligence Innovation and Incubation Institute, Fudan University\\
$^{2}$Shanghai Academy of AI for Science\\
$^{3}$FuXi Intelligent Computing Technology Co., Ltd.
}

\begin{document}

\maketitle

\begin{abstract}
Subseasonal precipitation forecasting is inherently uncertain due to chaotic atmospheric dynamics, making reliable uncertainty estimation essential for real-world applications. Existing approaches typically represent uncertainty through ensemble forecasts rather than directly modeling predictive distributions. However, due to systematic model biases, raw ensemble outputs are often not well calibrated and cannot be directly interpreted as reliable uncertainty estimates. As a result, operational systems rely on post-hoc calibration based on reforecast datasets, which are computationally expensive to generate and maintain.
To address these limitations, we propose \model, an end-to-end probabilistic forecasting framework with a dual-head design. The probabilistic and deterministic heads are supervised with separate objectives and optimized jointly. The framework further supports stochastic sampling, enabling probabilistic outputs even with a single stochastic forward pass and allowing optional multi-sample aggregation.
Extensive experiments show that \model\ demonstrates superior probabilistic forecasting skill while substantially reducing inference-time computational and storage costs.

\end{abstract}

\section{Introduction}
\label{intro}
\begin{wrapfigure}{r}{0.45\textwidth}
    \vspace{-10pt}
    \centering
    \includegraphics[width=0.43\textwidth]{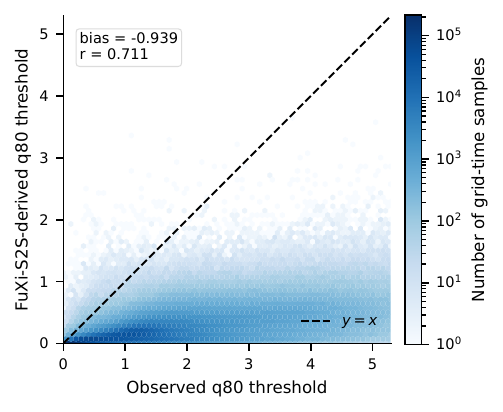}
    \vspace{-8pt}
    \caption{\small 
    Date-conditioned observed and model-derived (FuXi-S2S) q80 climatological thresholds at Week 3.
    Each sample corresponds to a grid cell and an initialization date.
    The systematic deviation below the diagonal indicates distribution mismatch in raw forecasts.
    }
    \label{fig:q80_mismatch}
    \vspace{-10pt}
\end{wrapfigure}

Forecasting total precipitation at the subseasonal scale, i.e., at lead times of roughly two to six weeks, is important for disaster preparedness, water management, agriculture, and energy systems~\cite{white2017potential,pegion2019subseasonal,bloomfield2021sub,white2022advances,domeisen2022advances}. Yet this regime is highly uncertain due to the chaotic nature of atmospheric evolution, where rapidly evolving weather systems interact with more slowly varying boundary forcings, leading to limited predictability~\cite{lorenz1979forced,mariotti2018progress,fuxi-s2s}. To characterize this uncertainty, operational centers such as ECMWF rely on ensemble prediction systems that sample uncertainties through perturbed initial conditions and model formulations~\cite{buizza2005comparison,leutbecher2008ensemble}, and recent machine-learning-based subseasonal systems have followed the same paradigm~\cite{fuxi-s2s,weyn2021sub,han2023ensemble,bach2024improved}. 

However, in most existing approaches, uncertainty is represented only implicitly through ensemble dispersion, while the training objective remains defined in the continuous value space through point-wise error minimization, rather than being directly aligned with the quantile structure of observed outcomes. As a result, raw ensemble forecasts are not explicitly constrained to provide reliable quantile-level interpretations. This issue is particularly severe for precipitation, whose distribution is highly skewed and heavy-tailed and whose forecasts often exhibit substantial model-specific distributional biases. Consequently, the same precipitation value can correspond to markedly different quantile levels under the forecast and observed distributions.
Figure~\ref{fig:q80_mismatch} illustrates this mismatch by comparing the 80th-percentile precipitation threshold derived from FuXi-S2S forecasts with that derived from observations. Although the two thresholds are positively correlated, the model-derived thresholds are systematically lower than their observed counterparts, with most samples lying below the one-to-one line. This indicates that raw ensemble distributions tend to underestimate the precipitation intensity associated with a given quantile level. Such a mismatch makes uncalibrated quantile interpretation unreliable and motivates post hoc calibration against historical reforecasts.
In practice, this calibration is especially costly for ensemble systems, since stable quantile estimation typically requires multi-member reforecasts over long historical periods, commonly around 20 years. This cost is recurrent rather than one-off: operational S2S systems are updated every one to two years, and each model cycle requires new reforecasts to estimate its model-specific biases and evaluate forecast skill consistently. As a result, reliable quantile interpretation remains tightly coupled with repeated reforecast generation, leading to substantial recurring computational and storage burdens.

Motivated by the limitations of post-calibration procedures, we propose \textbf{\model}, an end-to-end framework for direct subseasonal precipitation quantile forecasting. Built on a shared Swin Transformer backbone, \model\ uses a dual-head decoder to support autoregressive rollout while learning precipitation quantile distributions directly. To further improve probabilistic forecasting and reduce the mismatch between training and inference, we design an RPS-based loss function and an ensemble-consistent curriculum training strategy. Our contributions are outlined as follows:

\textbf{-- A paradigm shift from post-calibration to end-to-end quantile learning.}
We shift precipitation quantile estimation from post-processing to an end-to-end learning objective. \model\ uses a dual-head decoder, where an explicitly supervised regression branch supports stable autoregressive rollout, while the quantile branch directly predicts probability mass over climatological quantile bins derived from two decades of historical reanalysis data~\cite{hersbach2020era5}. This allows \model\ to learn quantile structure during training, eliminate \emph{the need for post hoc calibration} based on large multi-member reforecast archives, and support both single-member and ensemble-based forecasting with stochastic input perturbations~\cite{chen2024fuxiens}.

\textbf{-- An RPS-based probabilistic objective for distribution-aware quantile learning.} 
RPS evaluates the discrepancy between cumulative predicted and observed distributions. This allows the loss to account for the ordinal structure of climatological quantile bins and impose larger penalties on predictions farther from the observed category.

\textbf{-- An ensemble-consistent curriculum training strategy for training-inference alignment.} 
We first train \model\ on forecast steps 1 to 6. In the second training phase, ECCT extends training to forecast steps 12 to 18 and changes how samples are distributed across GPUs by grouping stochastic variants of the same forecast case on each GPU. This enables the quantile branch to learn from an ensemble-like predictive distribution during training, better aligns training with inference-time ensemble construction, and improves ensemble consistency.

\textbf{-- Empirical evidence.}
Extensive experimental results show that \model\ outperforms current state-of-the-art models in probabilistic forecasting skill while substantially reducing inference-time and storage costs for probabilistic forecasting.

\section{Related Work}
\label{sec:related_work}
\textbf{Deterministic Weather Forecasting.}
Deterministic forecasting has long been a dominant paradigm in weather prediction, with recent deep learning–based models achieving performance comparable to or surpassing traditional numerical approaches. Early work demonstrated efficient global forecasting via operator learning~\cite{pathak2022fourcastnet}, while subsequent models adopted more expressive architectures, such as transformer or graph neural network, and multi-step rollout strategies to improve long-range predictions~\cite{bi2023accurate,lam2023learning,chen2023fengwu,chen2023fuxi}. To further enhance model performance, recent studies have explored incorporating structural priors into neural forecasting models. Physics-informed approaches integrate conservation laws and domain knowledge into learning frameworks, improving physical consistency and long-term stability~\cite{kochkov2024neural,huang2025fuxi}, while geometry-aware models focus on better representing the spherical structure of the Earth system and spatial relationships in atmospheric data~\cite{cirt,liu2025equivariant}.
Despite these advances, deterministic forecasting remains fundamentally limited in representing the inherently chaotic nature of atmospheric dynamics and the associated predictive uncertainty. These limitations become especially pronounced at subseasonal (S2S) timescales, where uncertainty grows rapidly and small initial perturbations can lead to large deviations in forecast trajectories.

\textbf{Probabilistic Weather Forecasting.}
Probabilistic forecasting is another dominant paradigm in weather prediction, aiming to characterize the inherent uncertainty arising from chaotic atmospheric dynamics. 
Existing approaches can be broadly categorized into sampling-based methods and direct probabilistic forecasting. Sampling-based approaches represent uncertainty by generating multiple samples, forming ensembles that approximate the predictive distribution. Operational forecasting has long relied on ensemble prediction systems, where uncertainty is introduced via perturbations to initial conditions and model physics. Building on this paradigm, recent machine learning methods adopt similar strategies~\cite{price2024gencast,li2024seeds,fuxi-s2s,chen2024fuxiens} to model uncertainty implicitly through sampling. However, extracting explicit categorical probabilities or quantile estimates from such ensembles remains non-trivial. To correct systematic biases in raw ensembles, a large body of work focuses on postprocessing and calibration, typically supported by historical reforecast datasets that provide a basis for estimating model climatology and improving probabilistic reliability. Classical approaches such as ensemble model output statistics (EMOS)~\cite{gneiting2005calibrated}, Bayesian model averaging~\cite{raftery2005using}, and quantile regression forests~\cite{taillardat2016calibrated}, along with neural network–based extensions~\cite{rasp2018neural,bremnes2020ensemble, scheuerer2020using}, learn mappings from raw forecasts to calibrated probabilistic outputs. While effective, these methods operate as post-hoc steps and depend on ensemble or reforecast archives, limiting end-to-end probabilistic modeling.

Beyond these ensemble-based pipelines, a complementary class of approaches predicts probabilistic quantities directly from the model output. MetNet and its extension~\cite{sonderby2020metnet, espeholt2022deep} formulate precipitation forecasting as a classification problem over discretized intensity bins, directly producing categorical probabilities. However, such approaches are primarily developed for short-range and regional forecasting tasks and end-to-end subseasonal probabilistic forecasting remain underexplored in subseasonal settings. A full discussion of these works is included in Appendix~\ref{app_sec:related_work}

\section{Preliminaries}
\label{sec:preliminaries}

\paragraph{Problem Statement.}
We formulate subseasonal-to-seasonal (S2S) forecasting on a global latitude-longitude grid with a $1.5^\circ$ spatial resolution, covering $121 \times 240$ grid points over latitudes $[-90^\circ, 90^\circ]$ and longitudes $[-180^\circ, 180^\circ]$, with daily temporal resolution. The atmospheric state at time $t$ is represented as $\mathbf{X}^t \in \mathbb{R}^{C \times H \times W}$, where $C$, $H$, and $W$ denote the numbers of variables, latitudes, and longitudes, respectively. We consider $C{=}76$ variables, comprising 5 upper-air variables across 13 pressure levels and 11 surface variables, as detailed in Table~\ref{app_tab:variables}.

Given historical observations $\mathbf{X}^{t-T_1:t}$, the goal is to forecast the probabilistic state of future \textbf{total precipitation} as categorical probabilities over climatological quantile bins at subseasonal lead times:
{\small
\begin{equation}
    \mathbf{\hat{Q}}^{t+T_2:t+T_3} = f_\Theta(\mathbf{X}^{t-T_1:t}),
\end{equation}
}
where $\Theta$ denotes the neural network parameters, and $\mathbf{\hat{Q}}$ denotes the predicted categorical probability distribution over precipitation quantile bins. Following previous work~\cite{fuxi-s2s}, we set $T_1=2$, $T_2=15$, and $T_3=42$. Unlike previous S2S forecasting models that rely on ensemble regression forecasts $\mathbf{\hat{Y}}$ and large multi-member reforecast archives to obtain $\mathbf{\hat{Q}}$ through post-calibration and post-processing, our goal is to learn probabilistic forecasts directly within the model.

\textbf{Climatology Quintile Labels.}
Quintile labels are constructed from a climatological reference distribution based on a fixed 20-year period (2002–2021) of ERA5 data. 
To improve robustness against sampling variability and extreme events, daily values are defined as rolling weekly means computed from daily observations and aligned with the initialization time. 
Historical samples are then collected from calendar days at offsets of $\pm4$, $\pm2$, and 0 days relative to the target date, yielding 100 samples for each initialization date.
Based on these samples, empirical quintile thresholds (i.e., the 20th, 40th, 60th, and 80th percentiles) are computed independently for each spatial location, and target values are discretized into five categories according to these thresholds, which are used for both training and evaluation, following the setup in work~\cite{loegel2025ai}.

\textbf{Reforecast-based Calibration.}
Forecasting systems often have systematic biases, causing their raw prediction distributions to differ from the distribution of observations. 
Therefore, directly using observational climatology to define quantile bins can introduce substantial mismatch for model forecasts. 
Reforecasts mitigate this issue by providing historical forecast distributions for each initialization time and lead time. 
Instead of modifying the forecast values, this process uses the reforecast distribution to define model-specific quantile thresholds, and then discretizes the original forecasts based on these thresholds. 
In this paper, we refer to this reforecast-based quantile discretization process as calibration.

\section{Methodology}
\label{method}

\begin{figure}[t]
    \centering
    \includegraphics[width=\linewidth]{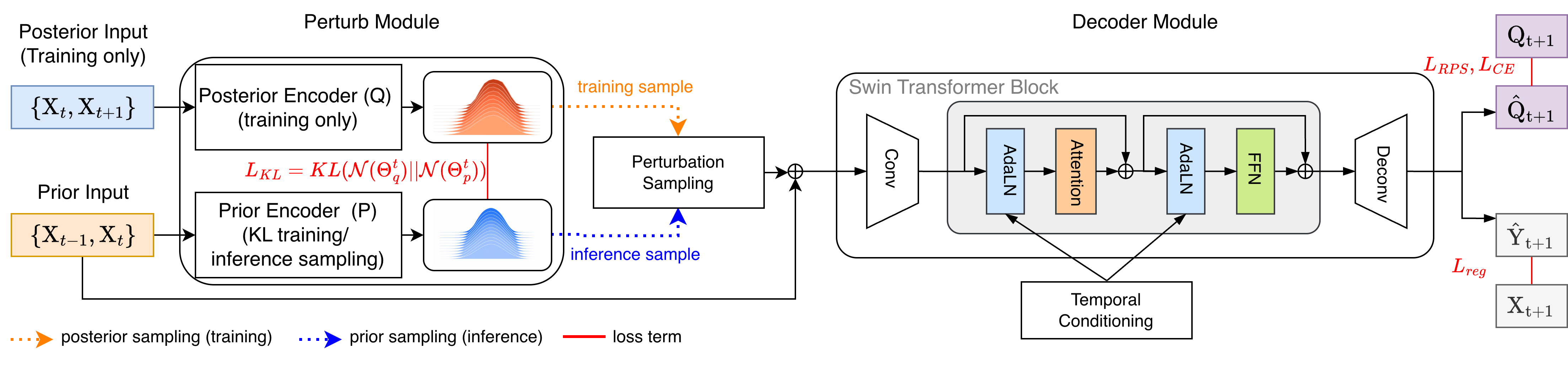}
    \caption{\small Architecture of \model\ framework. \model\ adopts a dual-head design, consisting of a regression head that produces regression forecast $\hat{\mathbf{Y}}^{t+1}$ for autoregressive rollout, and a probabilistic head that produces classification probabilities $\hat{\mathbf{Q}}^{t+1}$.  
    \underline{The Perturb Module} models stochastic variability. 
    The Perturb Module models stochastic variability via two encoders. 
    The posterior encoder $Q$ takes $\{\mathbf{X}^t, \mathbf{X}^{t+1}\}$ as input, while the prior encoder $P$ takes $\{\mathbf{X}^{t-1}, \mathbf{X}^t\}$. 
    A spatially correlated Gaussian perturbation is sampled and additively combined with the input state. {\color{orange}Orange:} posterior sampling path used for training; {\color{blue}blue:} prior distribution used for KL alignment during training and sampling during inference.
    The KL divergence loss $L_{\mathrm{KL}} = \mathrm{KL}(\mathcal{N}(\Theta_{q}^t)\,\|\, \mathcal{N}(\Theta_{p}^t))$ aligns the prior distribution with the posterior.
    \underline{The Decoder Module} processes the perturbed inputs using convolutional embedding followed by Swin Transformer blocks with Adaptive Layer Normalization, self-attention, and feed-forward networks with residual connections, modulated by temporal conditioning signals. 
    The model is jointly trained with $L_{\mathrm{KL}}$, $L_{reg}$, $L_{CE}$, and $L_{RPS}$ ({\color{red}red lines}).
    }
    \label{fig:framework}
\end{figure}

Figure~\ref{fig:framework} shows the overall architecture of our proposed \model, 
which builds upon the Swin Transformer Block and adopts an autoregressive forecasting paradigm. 
The model adopts a dual-head architecture, consisting of an auxiliary regression head and a probabilistic head. 
Under this formulation, S2S precipitation forecasting is cast as a joint optimization problem with an auxiliary autoregressive objective and a primary probabilistic objective. 
The regression head is explicitly supervised to produce continuous forecasts for subsequent rollout steps, while the probabilistic head directly predicts categorical probabilities over climatological quantile bins for total precipitation.
This enables \emph{direct probabilistic forecasting at inference time, eliminating the need for reforecast-based post-processing}.
We first detail the dual-head architecture in Section~\ref{subsec:dual-head}, 
which forms the core of our framework. We then introduce the training objectives for joint optimization in Section~\ref{subsec:joint_forecast} and present the perturbation mechanism for stochastic modeling in Section~\ref{subsec:perturb}. Finally, we illustrate our two-phase training strategy in Section~\ref{subsec:train_inference}.

\subsection{Dual-head Architecture}
\label{subsec:dual-head}
The model adopts a dual-head architecture built upon shared Swin Transformer blocks, followed by two task-specific output heads. 
Given the input states $\{\mathbf{X}^{t-1}, \mathbf{X}^{t}\}$, or their perturbed version, \model\ first encodes them with temporal conditioning embeddings into a shared latent representation $\mathbf{H}$. 
Meanwhile, a shallow feature $\mathbf{H}_0$ is extracted at the input stage and fused with $\mathbf{H}$ through a skip connection before output projection. 
The fused representation is then passed to two task-specific heads, each equipped with an independent adaptive layer normalization and output projection. 
The regression head produces a deterministic estimate $\hat{\mathbf{Y}}^{t+1}$ for autoregressive rollout, whereas the probabilistic head predicts classification probabilities $\hat{\mathbf{Q}}^{t+1}$ for probabilistic forecasting.

\textbf{Regression head.}
The regression branch forecasts the continuous field. It adopts a deconvolution-based projection that restores the original spatial resolution. To stabilize training, the prediction layers are zero-initialized, with separate branches for upper-atmosphere and surface variables.
{\small
\begin{equation}
    {\mathbf{\hat Y}}^{t+1} = f_{\text{reg}}(\mathbf{H}, \mathbf{H}_0).
\end{equation}
}

\textbf{Probabilistic head.}
In parallel, a classification branch models the relative position of the forecast within a reference distribution. 
It utilizes a projection layer that outputs $K$ logits per variable:
{\small
\begin{equation}
    \hat{\mathbf{Q}}^{t+1} = \frac{1}{\tau} \cdot f_{\text{cls}}(\mathbf{H}, \mathbf{H}_0),
\end{equation}
}
where $K$ is the number of quantile bins (set to 5 in this work), and $\tau$ is a learnable temperature parameter, which controls the sharpness of the categorical distribution. 
The probabilistic head is zero-initialized to avoid interfering with the regression branch during early training.

\subsection{Joint Learning}
\label{subsec:joint_forecast}
To jointly learn accurate point forecasts and predictive uncertainty, we optimize the dual-head model with a structured multi-objective training strategy. The regression head is supervised for point-wise accuracy to support autoregressive rollout, while the probabilistic head is trained to produce calibrated categorical distributions for probabilistic forecasting. 
A KL divergence term is further introduced to regularize stochastic perturbation learning, as detailed in Section~\ref{subsec:perturb}.

\textbf{Regression loss.}
The regression head is trained to produce accurate point forecasts to enable autoregressive rollout.
We adopt a latitude-weighted Charbonnier loss:
{\small
\begin{equation}
\mathcal{L}_{\mathrm{reg}}=
\frac{1}{C H W}
\sum_{c=1}^{C}\sum_{i=1}^{H}\sum_{j=1}^{W}
\alpha_i
\sqrt{
\left(\bar{Y}_{c,i,j}^{t+1}-X_{c,i,j}^{t+1}\right)^2+\epsilon^2
},
\end{equation}
}
Here, $\alpha_i = H \times \frac{\cos \Phi_i}{\sum_{i=1}^{H}\cos \Phi_i}$ is the latitude-dependent weighting factor at latitude $\Phi_i$. 
This weighting is applied to $\mathcal{L}_{\mathrm{RPS}}$ and $\mathcal{L}_{\mathrm{CE}}$ as well. 
The point prediction $\bar{Y}^{t+1}=\frac{1}{M}\sum_{m=1}^{M}\hat{Y}^{(m),t+1}$ is obtained as the mean of $M$ ensemble members. 

\textbf{Probabilistic loss.}
The probabilistic head predicts a categorical distribution over precipitation bins at each grid point. 
A natural supervision signal is the cross-entropy (CE) loss, which encourages the predicted probability mass to concentrate on the observed category:
{\small
\begin{equation}
\mathcal{L}_{\mathrm{CE}}=
\frac{1}{C \times H \times W}
\sum_{c=1}^{C}\sum_{i=1}^{H}\sum_{j=1}^{W}
\alpha_i
\left(
-\sum_{l=1}^{K}
Q^{t+1}_{c,i,j,l}
\log \hat{Q}^{t+1}_{c,i,j,l}
\right).
\end{equation}
}
Here, $\hat{Q}^{t+1}_{c,i,j,l}$ denotes the predicted categorical probability for bin $l$, and $Q^{t+1}_{c,i,j,l}$ denotes the corresponding one-hot encoded target. 
While CE provides direct supervision at the categorical level, it does not explicitly account for the ordinal distances between precipitation bins. 
As a result, misallocated probability mass is penalized mainly through the probability assigned to the observed bin, rather than according to how far the predicted mass is from the observed category.

To address this limitation, we further introduce the Ranked Probability Score (RPS), which evaluates probabilistic forecasts through their cumulative distributions:
{\small
\begin{equation}
\mathcal{L}_{\mathrm{RPS}}=
\frac{1}{C \times H \times W}
\sum_{c=1}^{C}\sum_{i=1}^{H}\sum_{j=1}^{W}
\alpha_i
\left(
\sum_{k=1}^{K}
\left(
\hat{Q}^{t+1}_{c,i,j}(k)
-
Q^{t+1}_{c,i,j}(k)
\right)^2
\right),
\end{equation}
}
where
\[
\hat{Q}^{t+1}_{c,i,j}(k)=\sum_{\ell=1}^{k}\hat{Q}^{t+1}_{c,i,j,\ell}, 
\quad
Q^{t+1}_{c,i,j}(k)=\sum_{\ell=1}^{k}Q^{t+1}_{c,i,j,\ell}.
\]
To avoid ambiguity, we use subscript notation to denote categorical probabilities and parentheses to denote cumulative distributions.

\subsection{Perturbation Module}
\label{subsec:perturb}
The probabilistic head enables end-to-end probabilistic forecasting without post-hoc calibration. 
We further introduce stochastic perturbations as an auxiliary mechanism to encourage forecast diversity and enrich the representation of uncertain spatial-temporal evolution. We apply perturbations in the input space rather than the latent space to preserve fine-scale spatial information, which is important for localized precipitation extremes. 

The perturbation module follows a variational formulation with two structurally identical branches, denoted as a prior network $P$ and a posterior network $Q$. At each forecast initialization time $t$, the prior network $P$ takes as input the past states $\{\mathbf{X}^{t-1}, \mathbf{X}^t\}$, while the posterior network $Q$ conditions on the future state $\{\mathbf{X}^{t},\mathbf{X}^{t+1}\}$. Both networks parameterize Gaussian distributions $\mathcal{N}(\Theta_{p}^t)$ and $\mathcal{N}(\Theta_{q}^t)$, respectively. To be mentioned, the posterior network only conditions on the immediate next state $\mathbf{X}_{t+1}$ during training, while all subsequent future states in the autoregressive rollout are generated by the model itself, without access to ground truth.

A perturbation field $\mathbf{z}^t$ is sampled from $\mathcal{N}(\Theta_{q}^t)$ during training and from $\mathcal{N}(\Theta_{p}^t)$ during inference, and is added to the input $\tilde{\mathbf{X}}^t = \mathbf{X}^t + \mathbf{z}^t$.
These perturbed inputs are then fed into the forecasting module to generate ensemble forecasts. The ensemble size is determined by the number of samples drawn from the prior distribution at inference time.

To stabilize the learning of stochastic perturbations, we introduce a KL divergence term that aligns the data-driven posterior with the generative prior:
{\small
\begin{equation}
\mathcal{L}_{\mathrm{KL}} =
\mathrm{KL}\left(\mathcal{N}(\Theta_{q}^{t}) \,\|\, \mathcal{N}(\Theta_{p}^{t})\right).
\end{equation}
}
The posterior distribution is used during training, whereas inference relies on the prior distribution.

\subsection{Training and Inference}
\label{subsec:train_inference}
\textbf{Model Training.}
\model\ is trained end-to-end with \emph{ensemble-consistent curriculum training (ECCT)}, 
a two-phase curriculum schedule that stabilizes autoregressive forecasting and aligns probabilistic 
supervision with ensemble-based inference. 
Since autoregressive prediction errors may accumulate over rollout steps, ECCT first trains the model 
with short-horizon rollouts, covering steps 1 to 6, and then extends the supervised rollout horizon 
to longer lead times, covering steps 12 to 18.
Throughout both phases, all model components are optimized jointly, with the same architecture and 
loss formulation. 
The overall training objective is defined as:
{\small
\begin{equation}
    \mathcal{L}
    =
    \mathcal{L}_\mathrm{reg}
    + \lambda_1 \, \mathcal{L}_\mathrm{RPS}
    + \lambda_2 \, \mathcal{L}_\mathrm{CE}
    + \lambda_3 \, \mathcal{L}_\mathrm{KL}.
\end{equation}
}

Beyond the rollout horizon, the two phases differ only in how samples are distributed across GPUs for 
probabilistic loss computation. 
In the first phase, each GPU processes different samples without broadcast, and the probabilistic 
losses are computed locally on each rank. 
In the second phase, the same sample is broadcast to all GPUs, allowing each GPU to generate an 
independent stochastic prediction for that sample. 
The probabilistic outputs are then gathered across GPUs and averaged to form an ensemble-mean 
probability, on which $\mathcal{L}_\mathrm{RPS}$ and $\mathcal{L}_\mathrm{CE}$ are computed. 
This does not change the model architecture or the loss formulation, but makes optimization consistent with ensemble-based probabilistic inference, where forecast probabilities are obtained by aggregating multiple stochastic predictions. In this sense, ECCT follows the autoregressive training recipe of FuXi-S2S, while adapting it to the end-to-end probabilistic setting through later-rollout supervision and ensemble-consistent probabilistic loss computation.

\textbf{Model Inference.}
During inference, \model\ samples perturbation fields from the prior perturbation model $P$ and additively fuses them with the input states to simulate uncertainty in the weather system. The perturbed inputs are then fed into the decoder with a dual-head structure, producing both deterministic forecasts from the regression head and probabilistic forecasts from the probabilistic head. Each sampling-and-forwarding pass generates one ensemble member, and repeating this process $M$ times produces $M$ members. We use the average output of the probabilistic head as the final uncertainty-aware prediction. By default, we set $M=8$ in our experiments. It is worth noting that even when $M=1$, \model\ can still produce probabilistic forecasts through its probabilistic head. 

\section{Experiments}
\label{sec:exp}

\subsection{Experimental Setup}
\textbf{Datasets.}
We use the ERA5 reanalysis dataset~\cite{hersbach2020era5} as the sole data source. ERA5 provides hourly global atmospheric variables from 1950 to present, which we aggregate into daily statistics on a $1.5^\circ$ grid (121 × 240). Data from 1979–2021 are used for training, and 2022 for evaluation. The climatological quantile labels and climatological reference forecasts are constructed from the fixed 20-year period 2002--2021, ensuring that no information from the test period is used.

\textbf{Competitors.}
\textbf{Climatology} is computed from a rolling 20-year historical window and used as the reference baseline for skill-based evaluation, ensuring that no information from the test period is used. 
We further compare \model\ with the operational ECMWF-S2S system and state-of-the-art learning-based model FuXi-S2S. 
\textbf{ECMWF-S2S} (denoted as EC-S2S) uses 11-member ensemble reforecasts from model cycle C47r3 for reforecast-based calibration, and 51-member real-time forecast outputs for evaluation over the same period. 
\textbf{FuXi-S2S} uses 51-member ensemble for both reforecast calibration and forecast evaluation, providing forecasts up to 42 days ahead. 
The evaluation focuses on subseasonal lead times from weeks 3 to 6, corresponding to lead days 15--42.

\textbf{Implementation Details.}
\model\ is implemented in PyTorch and optimized using AdamW. The loss weights are set to $\lambda_1=0.5$, $\lambda_2=0.1$, and $\lambda_3=5\times10^{-4}$, with a learning rate of $2\times10^{-4}$. The learnable temperature parameter in probabilistic header are set to $\tau=1$.
During curriculum training, each rollout step is trained for 1000 iterations. The model is trained on a cluster of 8 NVIDIA A100 GPUs, and inference is conducted on a single NVIDIA A100 GPU with 80\,GB memory.

\textbf{Evaluation Metrics}
\label{sec:metrics}
Following previous work~\cite{fuxi-s2s}, we evaluate probabilistic forecasting performance at both the grid-point and global levels, with global scores computed using latitude weighting over valid grid points. Specifically, we use the ranked probability skill score (RPSS) and Brier skill score (BSS) for probabilistic evaluation. For precipitation-based evaluation, extremely arid regions are excluded using the RPSS mask by default, while BSS uses a separate mask defined according to its binary-event evaluation setting. We also evaluate the computational efficiency and storage requirements of the proposed method. Although autoregressive rollout in the regression branch is not the primary learning objective, we report the anomaly root mean square error (RMSE), anomaly correlation coefficient (ACC), and temporal anomaly correlation coefficient (TCC) in Appendix~\ref{app_subsec:metrics}, same as other details.

\begin{figure}[!t]
    \centering
    \includegraphics[width=0.98\linewidth]{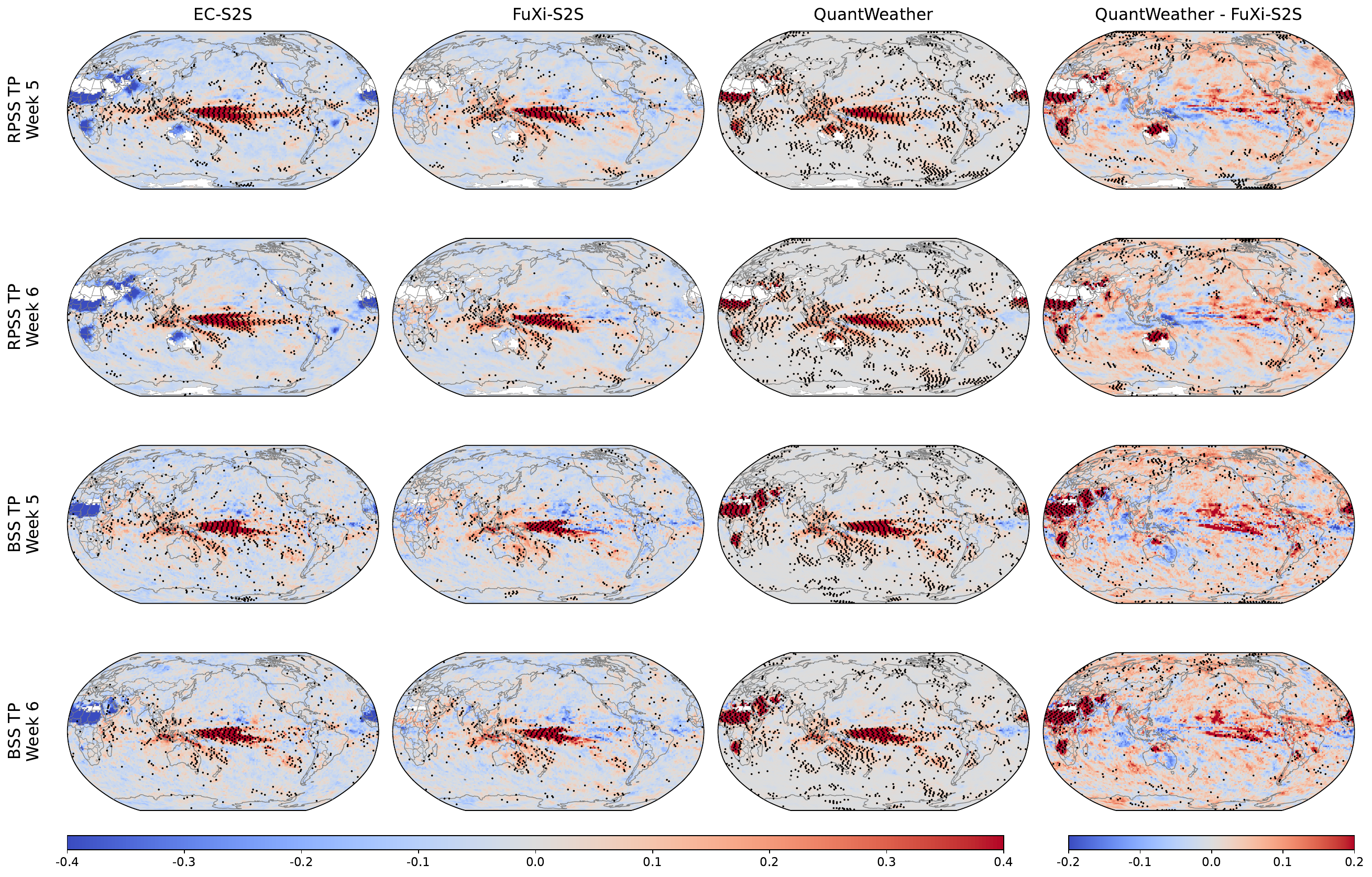}
    \caption{\small
    Average RPSS and BSS without latitude weighting for total precipitation (TP) at forecast lead times of weeks 5 and 6, evaluated using all testing data from 2022. Values closer to 1 indicate better skill for both metrics. The first three columns show ECMWF-S2S, FuXi-S2S, and \model, and the fourth column shows the difference between \model\ and the best baseline, FuXi-S2S. Red contour lines mark positive skill scores in the first three columns and positive differences in the fourth column. Stippling denotes statistical significance at the 97.5\% confidence level, indicating skill significantly above climatology in the first three columns and significant improvement over FuXi-S2S in the fourth column.
    }
    \label{fig:rpss_bss_global}
\end{figure}

\subsection{Main Results}

\textbf{Overall performance.} Figure~\ref{fig:rpss_bss_global} shows the spatial distributions of the temporally averaged RPSS and BSS for total precipitation. 
The RPSS evaluates the probabilistic skill across five climatological categories defined in Section~\ref{sec:preliminaries}, while the BSS focuses on extreme top 80th precipitation events. 
For both metrics, \model\ substantially reduces the regions with negative skill compared with all baselines, indicating more reliable probabilistic forecasts over a wider spatial extent. 
Moreover, \model\ achieves higher skill than FuXi-S2S in more regions, especially over land, where accurate precipitation forecasting is particularly important for practical applications.

\begin{figure}[!t]
    \centering
    \includegraphics[width=0.98\linewidth]{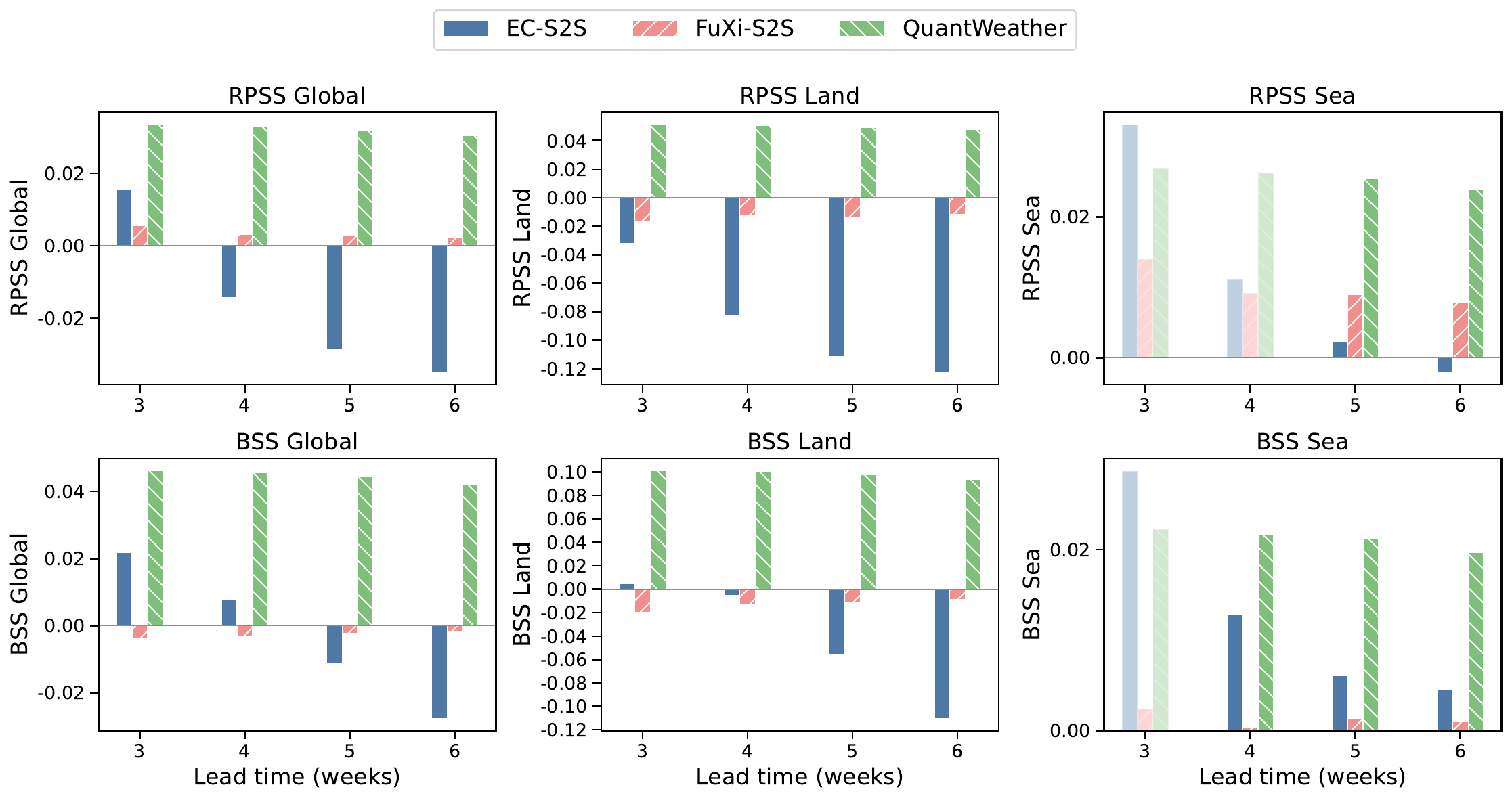}
    \vspace{-0.2cm}
    \caption{\small Comparison of latitude-weighted RPSS and BSS for TP forecasts from ECMWF-S2S, FuXi-S2S, and \model, evaluated using all testing data from 2022. Results are averaged over global, land, and sea regions, corresponding to the three columns. The two rows show RPSS and BSS, respectively, across forecast lead times from week 3 to week 6. Pale bars indicate cases where \model\ does not show a statistically significant improvement over FuXi-S2S at the 97.5\% confidence level.
    }
    \label{fig:avg_rpss_bss}
\end{figure}

Figure~\ref{fig:avg_rpss_bss} further summarizes the latitude-weighted global, land, and sea scores from week~3 to week~6. 
Overall, \model\ consistently achieves significantly higher RPSS and BSS than the baselines over the global and land regions during weeks~3--6. 
As the lead time increases, the performance of the baselines degrades more rapidly, whereas \model\ maintains relatively stable skill. 
This leads to an increasingly clear advantage of \model\ at longer lead times, suggesting that the proposed probabilistic forecasting framework provides more robust subseasonal precipitation predictions. 
For both figures, statistical significance is assessed using a bootstrapping approach repeated 1000 times.

\textbf{Model inference efficiency.} For the \emph{baseline models}, probabilistic forecasts are obtained from ensemble members. 
However, raw ensemble forecasts are affected by model-specific distributional biases, as shown in Figure~\ref{fig:q80_mismatch}. Therefore, post-hoc calibration based on reforecasts is required before these forecasts can be interpreted as reliable probabilistic predictions. Specifically, for each target initialization date, we construct model climatology from the corresponding calendar dates over the previous 20 years and use it to derive the required probabilistic references. Following previous work~\cite{fuxi-s2s}, the reforecast ensembles contain 11 members for ECMWF-S2S and 51 members for FuXi-S2S.
For real-time forecasting over the testing period, both baseline models use 51 ensemble members.
Table~\ref{tab:efficiency_comparison} reports the time and storage costs required for probabilistic evaluation. 
Unlike the baseline systems, \model\ does not require reforecast-based calibration and only performs inference over the testing year. 
It stores only the deterministic and probabilistic forecasts required for probabilistic evaluation, rather than maintaining a large multi-year reforecast archive. 
As a result, \model\ substantially reduces the storage cost and achieves an order-of-magnitude reduction in inference time compared with FuXi-S2S. 
These results demonstrate the efficiency advantage of the proposed end-to-end probabilistic forecasting framework at the inference stage.

\begin{table}[!t]
\centering
\caption{
\small Evaluation efficiency of S2S forecasting systems over one evaluation year with 104 initialization dates.
For reforecast-based systems, calibration uses a fixed 20-year reforecast archive.
Storage denotes the disk space required to store total-precipitation forecast outputs only.``s'' denotes second. ``N/A'' indicates that directly comparable measurements are not available.}
\label{tab:efficiency_comparison}
\begin{tabular}{l|ccccc}
\toprule
\multirow{2}{*}{\textbf{Model}} 
& \textbf{\# Members}
& \textbf{Reforecast}
& \textbf{Time} 
& \textbf{Total} 
& \textbf{Storage} \\
& {(reforecast/inference)}
& 
& {(s/init/member)} 
& {(s)} 
& {(GB)} \\
\midrule
ECMWF-S2S 
& 11/51
& \checkmark 
& N/A 
& N/A 
& 128.06 \\

FuXi-S2S 
& 51/51
& \checkmark 
& $\sim$14 
& $1.5 \times 10^{6}$ 
& 506.09 \\

\rowcolor{gray!10}
\textbf{Ours} 
& \textbf{-/8} 
& \textbf{\xmark} 
& $\sim$14 
& $\mathbf{1.2 \times 10^{4}}$ 
& $\mathbf{22.68}$ \\
\bottomrule
\end{tabular}
\vspace{0.3em}
\end{table}

\subsection{Ablation Study}
We compare \model\ with two variants: w/o-ECCT removes the second training phase, and w/o-RPS removes the RPS loss. 
Table~\ref{tab:ablation-week6} reports the Week-6 ablation results, where the effect of long-rollout probabilistic supervision is most pronounced. 
Removing ECCT consistently reduces both RPSS and BSS across global, land, and sea regions, indicating that ensemble-consistent curriculum training improves long-lead probabilistic forecasting. 
Removing RPS mainly affects RPSS, leading to less stable gains across lead times, while its effect on BSS is less consistent. This is expected because BSS evaluates threshold-based binary events, where the multi-bin precipitation distribution is collapsed into event and non-event probabilities. 
Therefore, the limitation of cross-entropy in not explicitly modeling the ordinal relationships among precipitation bins has a weaker and less direct effect on BSS than on RPSS, which evaluates the cumulative distribution over ordered categories. Full ablation results across Weeks 3--6 are reported in Table~\ref{app_tab:ablation} in Appendix~\ref{app_sec:experiments}.

\begin{wraptable}{r}{0.62\textwidth}
\vspace{-3.5em}
\centering
\caption{\small Ablation results of Week 6. ``$\uparrow$'' indicates larger is better. The best results are in \textbf{bold}.}
\label{tab:ablation-week6}
\small
\setlength{\tabcolsep}{3.5pt}
\begin{tabular}{lrrrrrr}
\toprule
\multirow{2}{*}{\textbf{Variant}} 
& \multicolumn{2}{c}{\textbf{Global}} 
& \multicolumn{2}{c}{\textbf{Land}} 
& \multicolumn{2}{c}{\textbf{Sea}} \\
\cmidrule(lr){2-3} \cmidrule(lr){4-5} \cmidrule(lr){6-7}
& RPSS$\uparrow$ & BSS$\uparrow$  & RPSS$\uparrow$  & BSS$\uparrow$  & RPSS$\uparrow$  & BSS$\uparrow$  \\
\midrule
w/o-ECCT & 0.025 & 0.040 & 0.045 & 0.092 & 0.018 & 0.017 \\
w/o-RPS  & 0.029 & \textbf{0.042} & \textbf{0.048} & \textbf{0.096} & 0.022 & 0.019 \\
\model   & \textbf{0.030} & \textbf{0.042} & \textbf{0.048} & 0.094 & \textbf{0.024} & \textbf{0.020} \\
\bottomrule
\end{tabular}
% \vspace{-1.2em}
\end{wraptable}

\subsection{Parameter Study}
We study the impact of the number of stochastic members used during inference, varying it over $\{1,2,4,8,16,32\}$. 
The results, reported in Figure~\ref{app_fig:parameter_study} in Appendix~\ref{app_sec:experiments}, show that the performance of \model\ improves slightly as the number of inference members increases. Notably, \model\ already achieves strong probabilistic performance with a single inference member, suggesting that the uncertainty modeling capability mainly benefits from the end-to-end dual-branch design rather than simply from increasing the ensemble size.

\textbf{Additional Results.} Appendix~\ref{app_sec:experiments} presents further results on model performance on other time periods, model performance versus number of ensemble members and regression results.

\section{Conclusion}
\label{sce:conclusion}
We shift subseasonal precipitation quantile estimation from a post-processing procedure to an end-to-end learning objective. To enable this paradigm, we propose \model, a dual-head framework that directly predicts categorical probabilities over climatological quantile bins while using an auxiliary regression branch to support stable autoregressive rollout. We further introduce an RPS-based probabilistic objective to better learn ordered precipitation distributions, and an ensemble-consistent curriculum training strategy to reduce the mismatch between training and inference.
We evaluate \model\ on the global ERA5 dataset for 2022. Experimental results show that \model\ consistently outperforms state-of-the-art models on probabilistic forecasting metrics while substantially reducing inference-time and storage costs for probabilistic forecasting. These results demonstrate the effectiveness of learning quantile structure directly within the forecasting model, rather than relying on post hoc calibration from large multi-member reforecast archives.

Although \model\ achieves strong probabilistic performance, introducing the probabilistic head may also affect the dispersion of stochastic samples. In our experiments, the single-member setting already provides competitive probabilistic skill, while increasing the number of inference members brings only modest additional gains. This suggests that the probabilistic branch effectively learns calibrated quantile information, but the joint optimization may also reduce the marginal benefit of ensemble diversity for the regression branch. 
Further exploring the balance between calibrated quantile prediction and ensemble dispersion is an important direction for future work.

\bibliographystyle{unsrt}
\bibliography{ref}
%%%%%%%%%%%%%%%%%%%%%%%%%%%%%%%%%%%%%%%%%%%%%%%%%%%%%%%%%%%%
%%%%%%%%%%%%%%%%%%%%%%%%%%%%%%%%%%%%%%%%%%%%%%%%%%%%%%%%%%%%

\newpage
\appendix
\section{Related Work}
\label{app_sec:related_work}
\subsection{Deterministic Weather Forecasting}
Deterministic forecasting has been one of the dominant paradigms in weather prediction, where recent deep learning–based models have demonstrated strong potential, achieving performance comparable to or even surpassing traditional numerical approaches. Early efforts such as FourCastNet~\cite{pathak2022fourcastnet} leveraged operator learning with Adaptive Fourier Neural Operators to produce fast global forecasts at 0.25° resolution. Building upon this, a series of models have advanced global spatial-temporal modelling through more expressive architectures. Transformer-based approaches such as Pangu-Weather~\cite{bi2023accurate} introduced 3D Earth-specific attention mechanisms and hierarchical temporal aggregation to improve long-range stability, while graph-based models such as GraphCast~\cite{lam2023learning} directly model atmospheric dynamics on multi-scale meshes and learn multi-step trajectories. Subsequent works further explored multi-modal learning, cascaded forecasting strategies, and large-scale pre-training to improve predictive skill and extend forecasting horizons~\cite{chen2023fengwu,chen2023fuxi,bodnar2024aurora}.

Beyond architectural advances, recent studies have explored incorporating structural priors into neural forecasting models. Physics-informed approaches integrate conservation laws and domain knowledge into learning frameworks, improving physical consistency and long-term stability~\cite{kochkov2024neural,huang2025fuxi}. Complementary to this line of work, geometry-aware models focus on better representing the spherical structure of the Earth system and spatial relationships in atmospheric data, which is particularly important for long-range and subseasonal (S2S) forecasting. Representative approaches include circular Transformer designs and equivariant message passing on the sphere that exploit spatial periodicity~\cite{cirt,liu2025equivariant}.

However, despite these advances, deterministic forecasting remains fundamentally limited in representing the inherently chaotic nature of atmospheric dynamics and the associated predictive uncertainty. These limitations become especially pronounced at subseasonal (S2S) timescales, where uncertainty grows rapidly and small initial perturbations can lead to large deviations in forecast trajectories.

\subsection{Probabilistic Weather Forecasting}
Probabilistic forecasting is another dominant paradigm in weather prediction, aiming to characterise the inherent uncertainty arising from chaotic atmospheric dynamics.
Existing approaches can be broadly categorised into sampling-based methods and direct probabilistic forecasting.

Sampling-based approaches represent uncertainty by generating multiple samples, forming an ensemble that approximates the predictive distribution. Operational weather forecasting has long relied on ensemble prediction systems, such as those developed by ECMWF, which generate multiple forecasts by perturbing initial conditions and model physics. Building on this paradigm, recent machine learning methods adopt similar strategies. Diffusion-based models, such as GenCast~\cite{price2024gencast} and SEEDS~\cite{li2024seeds}, generate diverse forecast trajectories through iterative denoising processes, while VAE-based approaches, such as FuXi-S2S~\cite{fuxi-s2s} and FuXi-ENS~\cite{chen2024fuxiens}, learn structured perturbations to efficiently produce ensemble members. While effective in capturing distributional variability, these methods represent uncertainty implicitly through samples and typically require multiple forward passes at inference time. Moreover, extracting explicit categorical probabilities or quantile estimates from such ensembles can be non-trivial.

To correct systematic biases in raw ensembles, a large body of work focuses on postprocessing and calibration, typically supported by historical reforecast datasets that provide a basis for estimating model climatology and improving probabilistic reliability. Classical approaches such as ensemble model output statistics (EMOS)~\cite{gneiting2005calibrated}, Bayesian model averaging~\cite{raftery2005using}, and quantile regression forests~\cite{taillardat2016calibrated}, as well as neural network–based postprocessing methods~\cite{rasp2018neural,bremnes2020ensemble}, convert deterministic or ensemble forecasts into calibrated probabilistic predictions. Related distributional regression methods~\cite{scheuerer2020using} parametrise predictive distributions and optimise likelihood-based or CRPS-based objectives. However, these approaches are applied \emph{after} model training and rely on the availability of ensemble forecasts or reforecast datasets, limiting their ability to fully exploit end-to-end representation learning.

Beyond sampling-based and postprocessing-based paradigms, a complementary class of approaches predicts probabilistic quantities directly from the model output. Direct probabilistic forecasting expresses uncertainty explicitly through quantities such as continuous quantile estimates or probabilities over predefined categories. For example, MetNet and its subsequent extension~\cite{sonderby2020metnet, espeholt2022deep} formulate precipitation forecasting as a classification problem over discretised intensity bins, directly producing categorical probabilities. However, such approaches are primarily developed for short-range and regional forecasting tasks, and remain underexplored in subseasonal probabilistic forecasting.

Overall, existing approaches predominantly rely on ensemble-based pipelines, where uncertainty is first represented through multiple samples and subsequently refined via postprocessing or calibration. While effective, such two-stage strategies do not produce probabilistic outputs directly from the model and are not naturally aligned with climatology-based categorical formats used in subseasonal forecasting. 

\section{Broader Impacts}
\label{app_sec:impacts}
\model\ represents a step toward a new paradigm for probabilistic subseasonal-to-seasonal forecasting.
Current S2S forecasting systems often rely on large ensembles and post-processing calibration to obtain usable probabilistic forecasts. 
While this pipeline has been effective, it usually requires substantial computational resources, forecast archives, and additional calibration procedures. 
By integrating probabilistic forecasting directly into the model training objective, \model\ moves uncertainty modeling from a post-hoc correction step to an end-to-end learning process. 
This provides a more direct and efficient way to generate calibrated probabilistic forecasts, and may reduce the dependence on expensive ensemble generation and separate post-calibration pipelines.

This paradigm can have broad implications for the development of next-generation S2S forecasting systems. 
Efficient end-to-end probabilistic forecasting may make subseasonal prediction more accessible to research groups, operational centers, and climate-service providers with limited computational and storage resources. 
It can also simplify the deployment of probabilistic forecast products, since uncertainty estimates are produced as part of the model output rather than being added through a separate calibration stage. 
For downstream applications, such as flood preparedness, drought monitoring, agricultural planning, water resource management, and disaster risk reduction, more accessible probabilistic forecasts can support earlier and better-informed preparation for high-impact precipitation events.

This new paradigm also opens up several directions for further study. 
Since learning-based probabilistic forecasts are trained on historical data, it remains important to examine their reliability under rare, extreme, or distribution-shifted conditions. 
Miscalibrated forecasts in these settings may mislead downstream users, and regional biases may persist when different climate regimes or extreme precipitation patterns are unevenly represented in the training data. 
In addition, although \model\ reduces the reliance on post-hoc calibration, continued forecast verification and reliability assessment remain important for practical deployment. 
These aspects provide promising directions for future work on robust and integrated probabilistic S2S forecasting.

\section{Additional Model Details}
Table~\ref{app_tab:variables} summarizes the input variables of \model. Table~\ref{app_tab:hyperparams} summarizes the key architectural choices and training hyperparameters.

\begin{table}[!t]
\centering
\caption{Summary of the upper-air and surface variables used as model inputs.}
\label{app_tab:variables}
\small
\begin{tabular}{l|l}
\toprule
Type & Full name \\
\midrule
Upper-air variables & geopotential \\
 & temperature \\
 & $u$ component of wind \\
 & $v$ component of wind \\
 & specific humidity \\
\midrule
Surface variables & 2\,m temperature \\
 & 2\,m dewpoint temperature \\
 & sea surface temperature \\
 & outgoing longwave radiation \\
 & 10\,m $u$ wind component \\
 & 10\,m $v$ wind component \\
 & 100\,m $u$ wind component \\
 & 100\,m $v$ wind component \\
 & mean sea-level pressure \\
 & total column water vapor \\
 & total precipitation \\
\bottomrule
\end{tabular}
\end{table}

\begin{table}[!t]
\centering
\caption{Architecture and training hyperparameters of \model.}
\label{app_tab:hyperparams}
\begin{tabular}{ll}
\toprule
\textbf{Component} & \textbf{Configuration} \\
\midrule
Resolution & 1.5\textdegree{} ($121 \times 240$ grid) \\
Channels & 76 (65 upper-atmosphere + 11 surface) \\
Patch size & $2 \times 2$ \\
Window size & $20 \times 20$ \\
Hidden dimension & 1536 \\
Attention heads & 24 \\
Decoder depth & $6 \times 6 = 36$ layers (Swin blocks) \\
Prior/posterior depth & $2 \times 6 = 12$ layers each \\
Attention type & Flash Attention \\
FFN type & SwiGLU \\
Normalisation & Adaptive Layer Norm (AdaLN) \\
Block type & Swin (shifted window) \\
Temporal embeddings & step, day-of-year (periodic) \\
\midrule
Quantile bins ($K$) & 5 (quintiles) \\
Classified variables & Total Precipitation (TP) \\
Temperature ($\tau$) & 1.0 (learnable) \\
\midrule
Latent channels ($C_z$) & $2 \times 76 = 152$ \\
Ensemble members & 8 (at inference) \\
\midrule
Optimiser & AdamW ($\text{lr} = 2\times10^{-4}$) \\
Precision & BF16 (params + reduce), FP32 (buffers) \\
Parallelism & FSDP2 with activation checkpointing \\
Rollout curriculum & 1--6 and 12--18 steps, +1 every 1{,}000 iters \\
Training period & 1979--2021 \\
Test period & 2022 \\
Frame interval & 24 h \\
\bottomrule
\end{tabular}
\end{table}

\section{Additional Experimental Details}
\label{app_sec:experiments}

\subsection{Evaluation Metrics}
\label{app_subsec:metrics}

We evaluate forecasting performance at both the grid-point and global levels. 
Global scores are computed using latitude weighting over valid grid points. 
We report the ranked probability skill score (RPSS) and Brier skill score (BSS) for probabilistic evaluation, and anomaly root mean square error (aRMSE), anomaly correlation coefficient (ACC), and temporal anomaly correlation coefficient (TCC) for deterministic evaluation. 
Let $s=1,\ldots,N$ index the evaluated forecast initializations, $t$ denote the lead time, and $(i,j)$ denote the grid point at latitude index $i$ and longitude index $j$, where $i=1,\ldots,H$ and $j=1,\ldots,W$. 
The latitude weight at latitude $\Phi_i$ is denoted as $\alpha_i$. 
For evaluations with spatial masks, we use $\mathbb{I}_{i,j}\in\{0,1\}$ to indicate whether grid point $(i,j)$ is valid.

\paragraph{Anomaly fields.}
For deterministic evaluation, anomalies for all variables are defined as deviations from the climatological mean calculated over the 20-year period from 2002 to 2021, denoted as $\mathbf{X}'$ and $\hat{\mathbf{Y}}'$ for observations and forecasts, respectively.

\paragraph{aRMSE.}
The anomaly root mean square error at lead time $t$ is computed as
\begin{equation}
\mathrm{aRMSE}_{t}
=
\sqrt{
\frac{
\sum_{i=1}^{H}\sum_{j=1}^{W}
\mathbb{I}_{i,j}\alpha_i
\left(
\frac{1}{N}
\sum_{s=1}^{N}
\left(
\hat{Y}'_{s,t,i,j} - X'_{s,t,i,j}
\right)^2
\right)
}{
\sum_{i=1}^{H}\sum_{j=1}^{W}
\mathbb{I}_{i,j}\alpha_i
}
},
\end{equation}
Lower aRMSE indicates better forecasting performance.

\paragraph{ACC.}
The anomaly correlation coefficient measures the spatial correlation between forecast and observed anomaly fields. 
For each forecast initialization $s$ and lead time $t$, ACC is computed as
\begin{equation}
\mathrm{ACC}_{s,t}
=
\frac{
\sum_{i=1}^{H}\sum_{j=1}^{W}
\mathbb{I}_{i,j}\alpha_i
\hat{Y}'_{s,t,i,j}
X'_{s,t,i,j}
}{
\sqrt{
\left(
\sum_{i=1}^{H}\sum_{j=1}^{W}
\mathbb{I}_{i,j}\alpha_i
\left(\hat{Y}'_{s,t,i,j}\right)^2
\right)
\left(
\sum_{i=1}^{H}\sum_{j=1}^{W}
\mathbb{I}_{i,j}\alpha_i
\left(X'_{s,t,i,j}\right)^2
\right)
}
}.
\end{equation}
The reported ACC at lead time $t$ is averaged over all evaluated forecast initializations:
\begin{equation}
\overline{\mathrm{ACC}}_{t}
=
\frac{1}{N}
\sum_{s=1}^{N}
\mathrm{ACC}_{s,t}.
\end{equation}
Higher ACC indicates better agreement between forecast and observed anomaly patterns.

\paragraph{TCC.}
The temporal anomaly correlation coefficient measures the temporal correlation between forecast and observed anomaly time series at each grid point. 
For each lead time $t$ and grid point $(i,j)$, TCC is computed as
\begin{equation}
\mathrm{TCC}_{t,i,j}
=
\frac{
\sum_{s=1}^{N}
\hat{Y}'_{s,t,i,j}
X'_{s,t,i,j}
}{
\sqrt{
\left(
\sum_{s=1}^{N}
\left(\hat{Y}'_{s,t,i,j}\right)^2
\right)
\left(
\sum_{s=1}^{N}
\left(X'_{s,t,i,j}\right)^2
\right)
}
}.
\end{equation}
The global TCC at lead time $t$ is obtained by latitude-weighted averaging over valid grid points:
\begin{equation}
\overline{\mathrm{TCC}}_{t}
=
\frac{
\sum_{i=1}^{H}\sum_{j=1}^{W}
\mathbb{I}_{i,j}\alpha_i
\mathrm{TCC}_{t,i,j}
}{
\sum_{i=1}^{H}\sum_{j=1}^{W}
\mathbb{I}_{i,j}\alpha_i
}.
\end{equation}
Higher TCC indicates better temporal consistency of the anomaly forecasts.

\paragraph{RPS and RPSS.}
For categorical probabilistic forecasts with $K$ ordered bins, let 
$p_{s,t,i,j,k}$ denote the predicted probability for bin $k$, and let 
$q_{s,t,i,j,k}$ denote the corresponding one-hot encoded observation, where
$s=1,\ldots,N$ indexes the evaluated forecast initializations, $t$ is the lead time, and $(i,j)$ denotes the grid point. 
The Ranked Probability Score at forecast initialization $s$, lead time $t$, and grid point $(i,j)$ is defined as
\begin{equation}
\mathrm{RPS}_{s,t,i,j}
=
\sum_{k=1}^{K}
\left(
\sum_{\ell=1}^{k} p_{s,t,i,j,\ell}
-
\sum_{\ell=1}^{k} q_{s,t,i,j,\ell}
\right)^2 .
\end{equation}
The latitude-weighted RPS at lead time $t$ is computed over valid grid points as
\begin{equation}
\overline{\mathrm{RPS}}_t
=
\frac{
\sum_{i=1}^{H}\sum_{j=1}^{W}
\mathbb{I}^{\mathrm{RPSS}}_{i,j}\alpha_i
\left(
\frac{1}{N}
\sum_{s=1}^{N}
\mathrm{RPS}_{s,t,i,j}
\right)
}{
\sum_{i=1}^{H}\sum_{j=1}^{W}
\mathbb{I}^{\mathrm{RPSS}}_{i,j}\alpha_i
},
\end{equation}
where $\mathbb{I}^{\mathrm{RPSS}}_{i,j}\in\{0,1\}$ denotes the valid mask used for RPSS evaluation, and $\alpha_i$ is the latitude-dependent weight. 
The Ranked Probability Skill Score is then computed relative to the climatological probabilistic forecast:
\begin{equation}
\mathrm{RPSS}_t
=
1
-
\frac{
\overline{\mathrm{RPS}}^{\mathrm{model}}_t
}{
\overline{\mathrm{RPS}}^{\mathrm{clim}}_t
}.
\end{equation}
Here, $\overline{\mathrm{RPS}}^{\mathrm{clim}}_t$ is computed using the climatological categorical forecast. 
Since the precipitation bins are defined by quantiles estimated from historical observations over the training period, the climatological probability is approximately uniform across bins, i.e., $p^{\mathrm{clim}}_{t,i,j,k}\approx 1/K$.

\paragraph{BS and BSS.}
For binary events, such as precipitation exceeding a predefined threshold, let 
$\hat{p}^{(e)}_{s,t,i,j}$ denote the predicted event probability and let 
$q^{(e)}_{s,t,i,j}\in\{0,1\}$ denote the observed event label. 
The Brier Score at forecast initialization $s$, lead time $t$, and grid point $(i,j)$ is defined as
\begin{equation}
\mathrm{BS}_{s,t,i,j}
=
\left(
\hat{p}^{(e)}_{s,t,i,j}
-
q^{(e)}_{s,t,i,j}
\right)^2 .
\end{equation}
The latitude-weighted BS at lead time $t$ is computed over valid grid points as
\begin{equation}
\overline{\mathrm{BS}}_t
=
\frac{
\sum_{i=1}^{H}\sum_{j=1}^{W}
\mathbb{I}^{\mathrm{BSS}}_{i,j}\alpha_i
\left(
\frac{1}{N}
\sum_{s=1}^{N}
\mathrm{BS}_{s,t,i,j}
\right)
}{
\sum_{i=1}^{H}\sum_{j=1}^{W}
\mathbb{I}^{\mathrm{BSS}}_{i,j}\alpha_i
},
\end{equation}
where $\mathbb{I}^{\mathrm{BSS}}_{i,j}\in\{0,1\}$ denotes the valid mask used for BSS evaluation. 
The Brier Skill Score is then computed relative to the climatological probabilistic forecast:
\begin{equation}
\mathrm{BSS}_t
=
1
-
\frac{
\overline{\mathrm{BS}}^{\mathrm{model}}_t
}{
\overline{\mathrm{BS}}^{\mathrm{clim}}_t
}.
\end{equation}
Here, $\overline{\mathrm{BS}}^{\mathrm{clim}}_t$ is computed using the climatological event probability. 
For events defined by the $\tau$-quantile threshold estimated from historical observations, the climatological event probability corresponds to the empirical exceedance frequency and is approximately $1-\tau$.

\subsection{Asset Licenses and Terms}
\label{app_subsec:liceses}
We use ERA5 reanalysis data as the primary data source. ERA5 is publicly available through the Copernicus Climate Data Store and is used in accordance with the Copernicus data access terms and license. 
We use ECMWF-S2S forecast and reforecast products for baseline comparison and evaluation, following the corresponding ECMWF data access terms. 
We also compare with FuXi-S2S using publicly described model outputs and evaluation protocols from prior work, with the original creators properly cited. 
All external datasets, models, and baselines used in this paper are credited through their original publications or data providers, and no proprietary or personally identifiable data are used.

\subsection{Reproducibility Details}
\label{app_subsec:implement}
\paragraph{Data periods and climatology construction.}
All experiments use ERA5 as the data source. 
The model training period is 1979--2021, and the test period is 2022. 
The climatological quantile thresholds used for label construction and skill-score references are computed from the fixed 20-year period 2002--2021. 
This separation ensures that the evaluation year is not used for either model training or climatological reference construction. 
For each target calendar date, historical samples are collected from nearby calendar days within the reference period, following the climatological label construction described in Section~\ref{sec:preliminaries}.

\paragraph{Evaluation initialization dates.}
Following the ECMWF-S2S evaluation protocol, forecasts are initialized twice per week, on Mondays and Thursdays. 
For the 2022 evaluation period, this gives 104 initialization dates. 
All models are evaluated on the same set of initialization dates and lead times from Week 3 to Week 6, corresponding to lead days 15--42.

\paragraph{Arid-region masks.}
Following previous work, we exclude extremely dry regions using static masks derived from climatological precipitation quantile thresholds.
The mask value is 1 for retained grid points and 0 for masked grid points. 
For RPSS, dry regions are identified using the lowest quintile boundary, while for BSS, which evaluates upper-quintile precipitation events, they are identified using the highest quintile boundary. 
Grid points with the corresponding climatological threshold no larger than 0.005 mm are excluded.
In implementation, date-dependent candidate masks are first constructed from the climatological thresholds and aligned with the valid calendar dates of the evaluated forecasts. 
These candidate masks are then aggregated over the evaluated initialization dates and target lead weeks to obtain one static mask for each metric. 
The same static mask is applied to all models for a given metric, ensuring fair comparison over identical retained grid points.

\begin{figure}[!t]
    \centering
    \includegraphics[width=\linewidth]{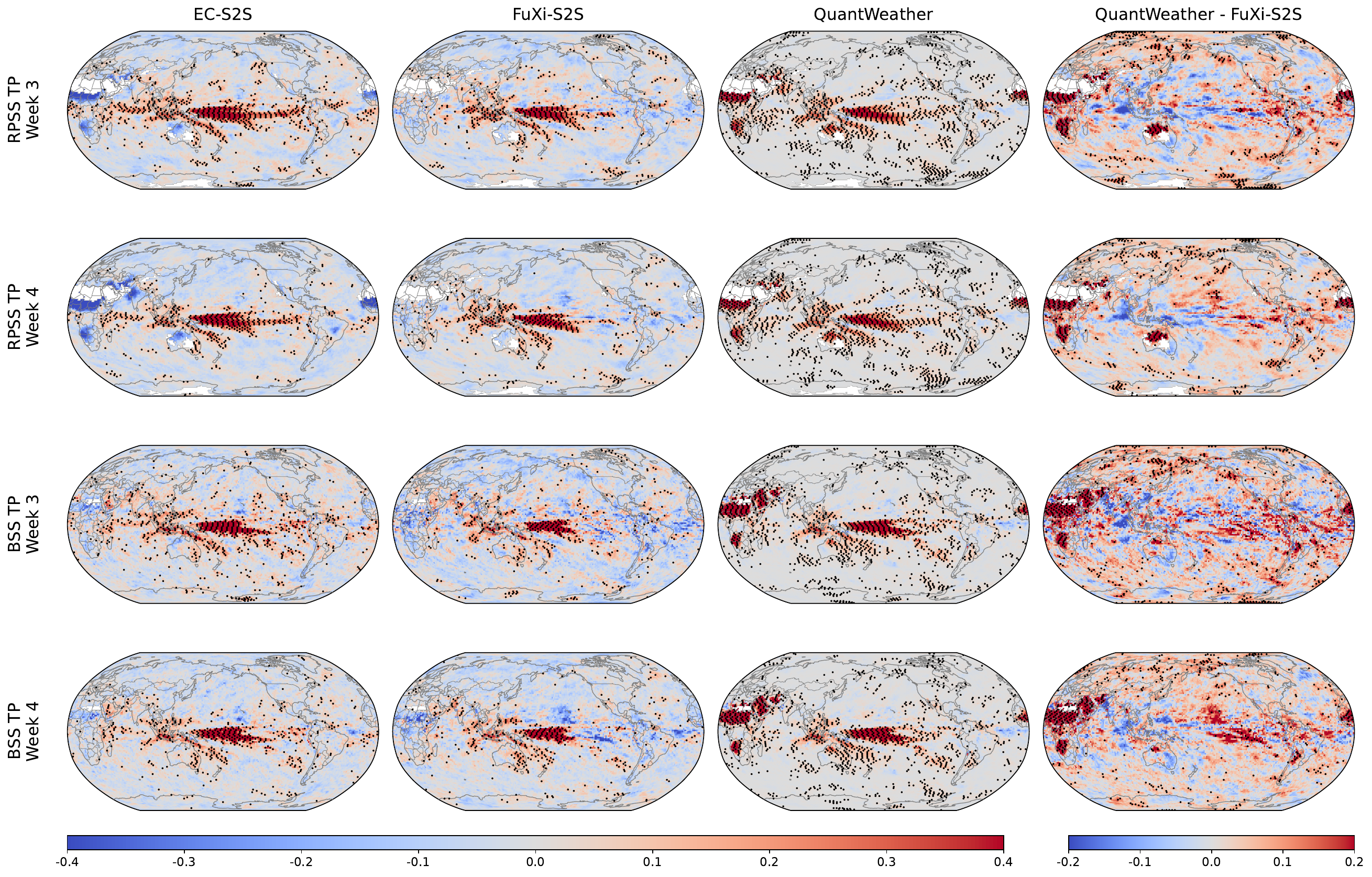}
    \caption{\small
    Average RPSS without latitude weighting for total precipitation (TP) at forecast lead times of week 3 and week4, evaluated using all testing data from 2022. The first three columns show the RPSS of the baselines and \model, and the fourth column shows the RPSS difference between \model\ and the best baseline, FuXi-S2S. Red contour lines mark positive RPSS in the first three columns and positive differences in the fourth column. Stippling denotes statistical significance at the 97.5\% confidence level. In the first three columns, stippling indicates skill significantly above climatology, while in the fourth column, it indicates that \model\ significantly outperforms FuXi-S2S.
    }
    \label{app_fig:rpss_bss_global}
\end{figure}

\subsection{Main Results}
\paragraph{Overall performance.} Figure~\ref{app_fig:rpss_bss_global} shows the spatial distributions of the temporally averaged RPSS and BSS for total precipitation of Week 3 and Week 4. 
The RPSS evaluates the probabilistic skill across five climatological categories defined in Section~\ref{sec:preliminaries}, while the BSS focuses on extreme top 80th precipitation events. 
For both metrics, \model\ substantially reduces the regions with negative skill compared with all baselines, indicating more reliable probabilistic forecasts over a wider spatial extent. 
Moreover, \model\ achieves higher skill than FuXi-S2S in more regions, especially over land, where accurate precipitation forecasting is particularly important for practical applications.

\begin{figure}[!t]
    \centering
    \includegraphics[width=0.98\linewidth]{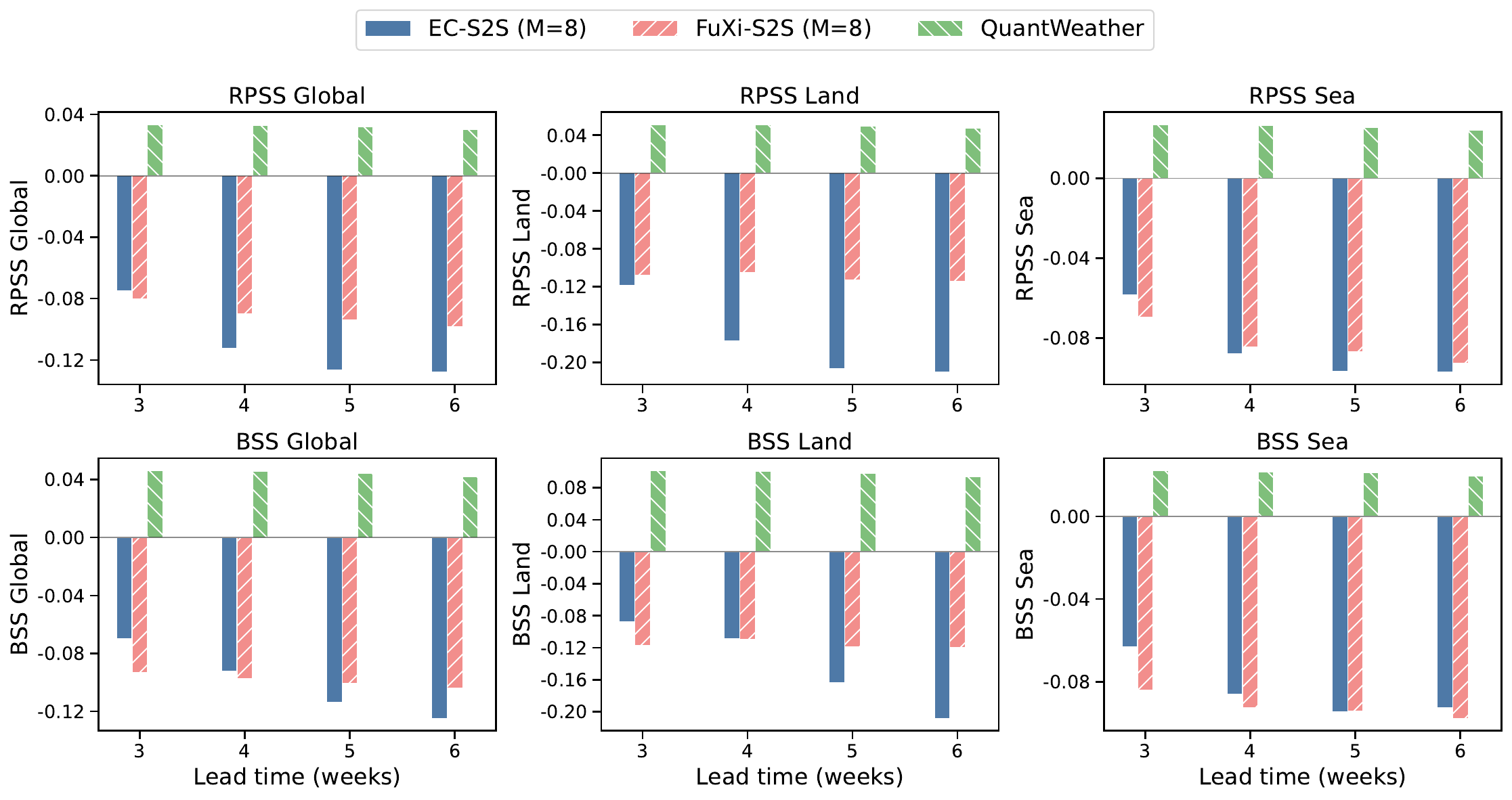}
    \caption{\small Comparison of latitude-weighted RPSS and BSS for TP forecasts from EC-S2S, FuXi-S2S, and \model with same number of ensemble member (M=8), evaluated using all testing data from 2022. Results are averaged over global, land, and sea regions, corresponding to the three columns. The two rows show RPSS and BSS, respectively, across forecast lead times from week 3 to week 6. Pale bars indicate cases where \model\ does not show a statistically significant improvement over FuXi-S2S at the 97.5\% confidence level.
    }
    \label{app_fig:avg_rpss_bss}
\end{figure}

\paragraph{Comparison with the same number of inference members.}
The main results compare \model\ with 8 stochastic members against EC-S2S and FuXi-S2S with 51 ensemble members, where \model\ already achieves better RPSS and BSS. 
To further ensure a fair comparison in terms of inference ensemble size, we evaluate all methods using 8 members. 
Figure~\ref{app_fig:avg_rpss_bss} shows that the baseline models obtain mostly negative RPSS and BSS, whereas \model\ maintains positive skill scores. 
This clear gap indicates that simply reducing the ensemble size of conventional forecasts greatly weakens their probabilistic skill, while \model\ remains effective with a small number of inference members. 
These results suggest that \model\ provides stronger uncertainty estimation with fewer stochastic samples, rather than relying on a large ensemble to obtain probabilistic skill.

\subsection{Ablation Study}
We compare \model\ with two variants: w/o-ECCT removes the second training phase, and w/o-RPS removes the RPS loss. Table~\ref{app_tab:ablation} reports the Week 3 to Week 6 ablation results, where the effect of long-rollout probabilistic supervision is most pronounced. 

\begin{table}[t]
    \centering
    \caption{Ablation Study from Week 3 to Week 6}
    \label{app_tab:ablation}
    \begin{tabular}{l|l|l|rrrr}
\toprule
Metric & Region & Model & Week 3 & Week 4 & Week 5 & Week 6 \\
\midrule
\multirow{9}{*}{RPSS}
& \multirow{3}{*}{Global}
& QuantWeather-w/o-ECCT & 0.027 & 0.027 & 0.026 & 0.025 \\
& & QuantWeather-w/o-RPS   & \textbf{0.033} & 0.032 & 0.031 & 0.029 \\
& & QuantWeather           & \textbf{0.033} &\textbf{ 0.033} & \textbf{0.032} &\textbf{ 0.030} \\
\cmidrule(lr){2-7}
& \multirow{3}{*}{Land}
& QuantWeather-w/o-ECCT & 0.047 & 0.047 & 0.046 & 0.045 \\
& & QuantWeather-w/o-RPS   & \textbf{0.052} & \textbf{0.051} & \textbf{0.050} &\textbf{0.048} \\
& & QuantWeather           & 0.051 & \textbf{0.051} & 0.049 & \textbf{0.048} \\
\cmidrule(lr){2-7}
& \multirow{3}{*}{Sea}
& QuantWeather-w/o-ECCT & 0.020 & 0.020 & 0.019 & 0.018 \\
& & QuantWeather-w/o-RPS   & 0.026 & 0.025 & 0.024 & 0.022 \\
& & QuantWeather           &\textbf{ 0.027} & \textbf{0.026} & \textbf{0.025} & \textbf{0.024} \\
\midrule
\multirow{9}{*}{BSS}
& \multirow{3}{*}{Global}
& QuantWeather-w/o-ECCT & 0.042 & 0.042 & 0.041 & 0.040 \\
& & QuantWeather-w/o-RPS   & \textbf{0.046} & \textbf{0.046} & \textbf{0.044} &\textbf{ 0.042} \\
& & QuantWeather           & \textbf{0.046} & \textbf{0.046} & \textbf{0.044} & \textbf{0.042} \\
\cmidrule(lr){2-7}
& \multirow{3}{*}{Land}
& QuantWeather-w/o-ECCT & 0.096 & 0.096 & 0.094 & 0.092 \\
& & QuantWeather-w/o-RPS   & \textbf{0.102} & \textbf{0.101} & \textbf{0.099} & \textbf{0.096} \\
& & QuantWeather           & 0.101 & \textbf{0.101} & 0.098 & 0.094 \\
\cmidrule(lr){2-7}
& \multirow{3}{*}{Sea}
& QuantWeather-w/o-ECCT & 0.018 & 0.018 & 0.018 & 0.017 \\
& & QuantWeather-w/o-RPS   & \textbf{0.022} & \textbf{0.021} & 0.020 & 0.019 \\
& & QuantWeather           & \textbf{0.022} & \textbf{0.022} & \textbf{0.021} & \textbf{0.020} \\
\bottomrule
\end{tabular}
\end{table}

\subsection{Parameter Study}
\begin{figure}[t]
    \centering
    \includegraphics[width=\linewidth]{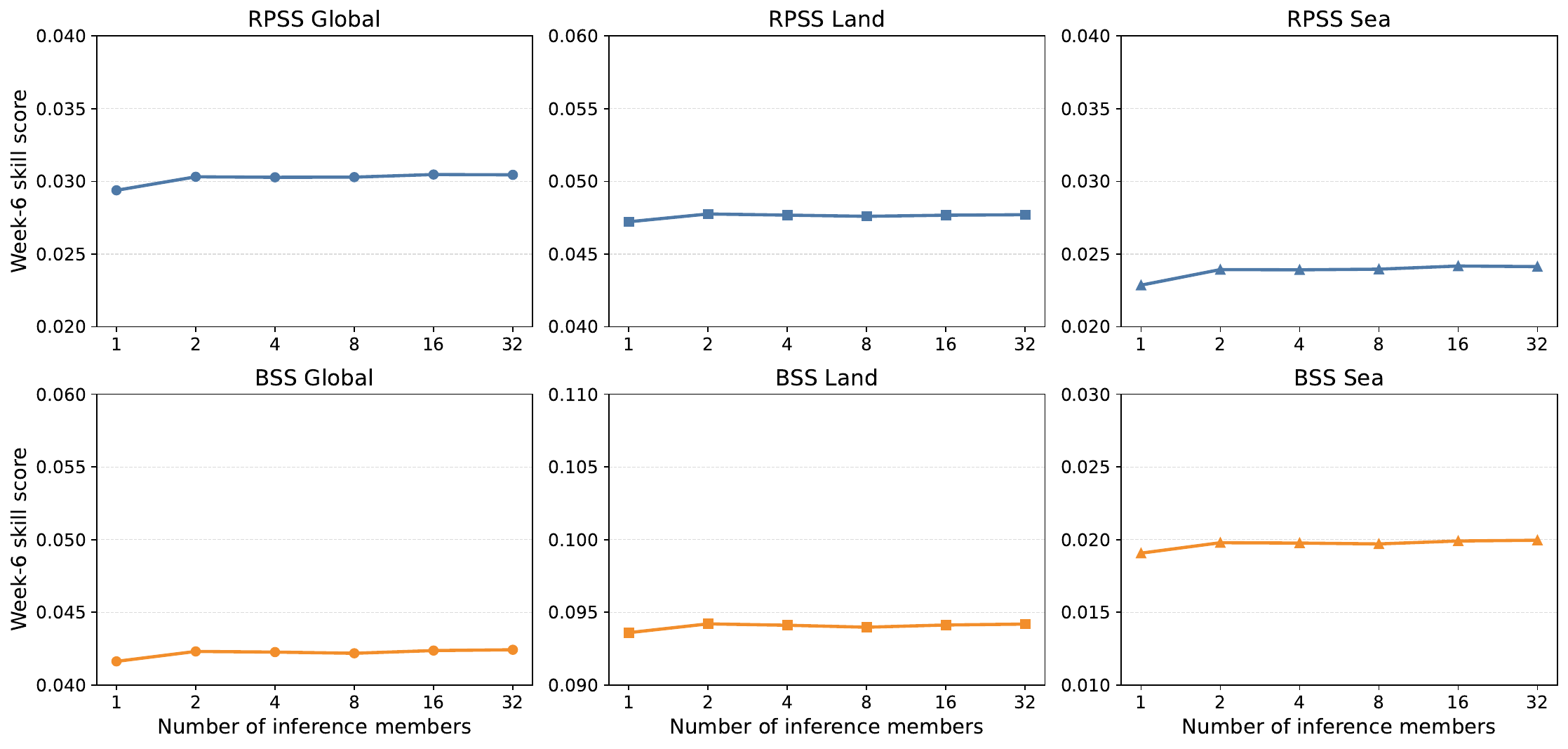}
    \caption{\small
    Parameter study. Varying number of ensemble member from 1 to 32. Report latitude-averaged RPSS and BSS total precipitation (TP) at forecast lead times of week 6, evaluated using all testing data from 2022.
    }
    \label{app_fig:parameter_study}
\end{figure}

We study the impact of the number of stochastic members used during inference, varying it over $\{1,2,4,8,16,32\}$. 
Figure~\ref{app_fig:parameter_study} shows that the performance of \model\ improves slightly as the number of inference members increases. Notably, \model\ already achieves strong probabilistic performance with a single inference member, suggesting that the uncertainty modeling capability mainly benefits from the end-to-end dual-branch design rather than simply from increasing the ensemble size.

\subsection{Regression Results}
We further evaluate the regression branch under different inference ensemble sizes. 
When using a single member ($M=1$), \model\ outperforms the baselines in most cases, achieving better results in 17 out of 24 comparisons. 
This indicates that introducing the probabilistic branch does not degrade the quality of the regression branch. 
Instead, the joint optimization can even provide a slight improvement to deterministic forecasting.

However, when the number of inference members increases to $M=8$, the improvement of \model\ on regression metrics becomes less pronounced. 
In contrast, the two baseline models benefit more clearly from the enlarged ensemble size. 
This suggests that although the probabilistic branch enables stronger probabilistic forecasting than conventional ensemble-based methods, it may also constrain the dispersion of sampled members. 
As a result, the potential gain of the regression branch from ensemble averaging is partially limited.

This suggests a possible trade-off introduced by the probabilistic branch. 
While it improves probabilistic forecasting by learning a more calibrated predictive distribution, it may also reduce the diversity among sampled members, thereby limiting the additional benefit that the regression branch can obtain from ensemble averaging.

\begin{figure}[t]
    \centering

    \begin{subfigure}[t]{0.98\linewidth}
        \centering
        \includegraphics[width=\linewidth]{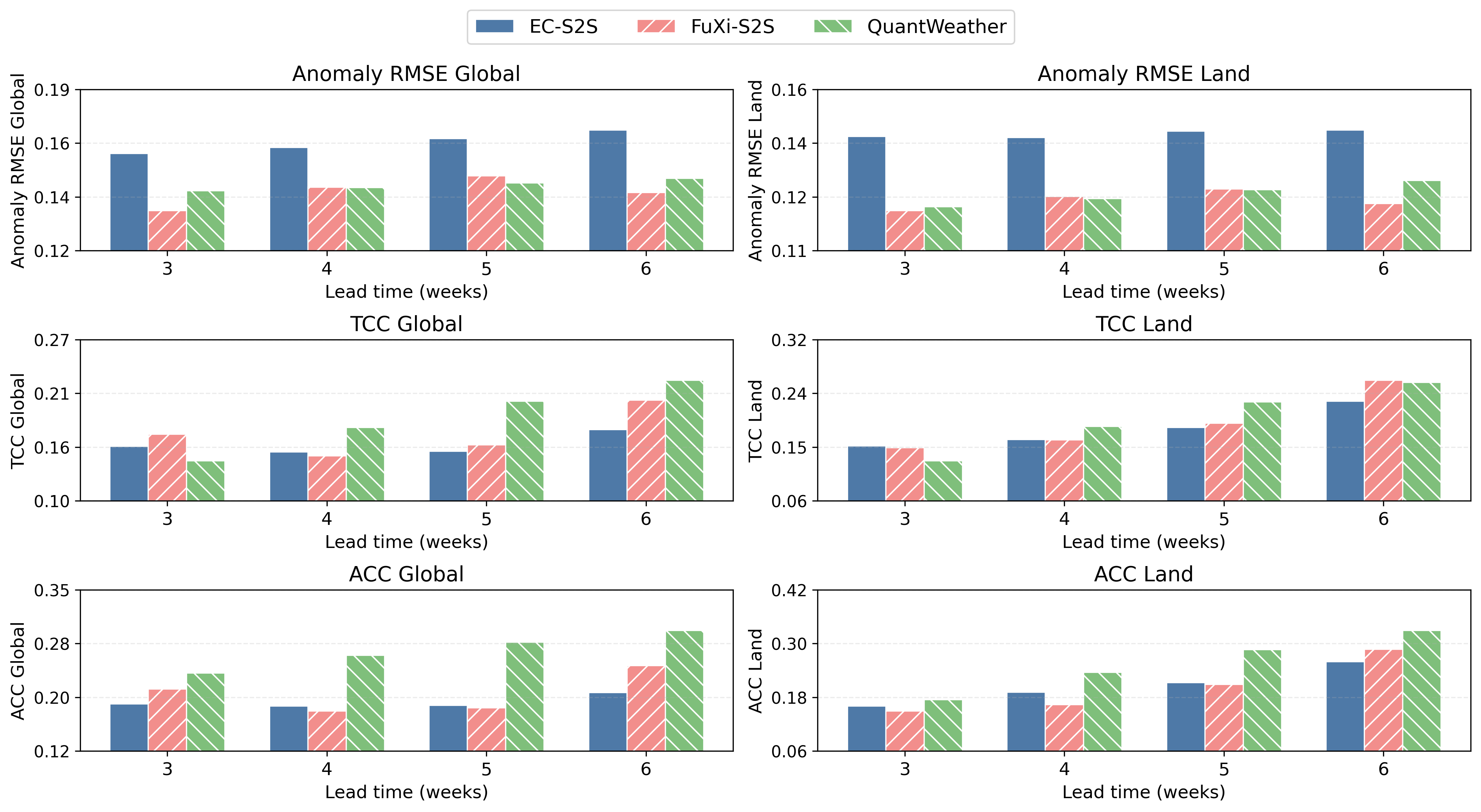}
        \caption{Member = 1}
        \label{app_fig:reg_m1}
    \end{subfigure}
    \hfill
    \begin{subfigure}[t]{0.98\linewidth}
        \centering
        \includegraphics[width=\linewidth]{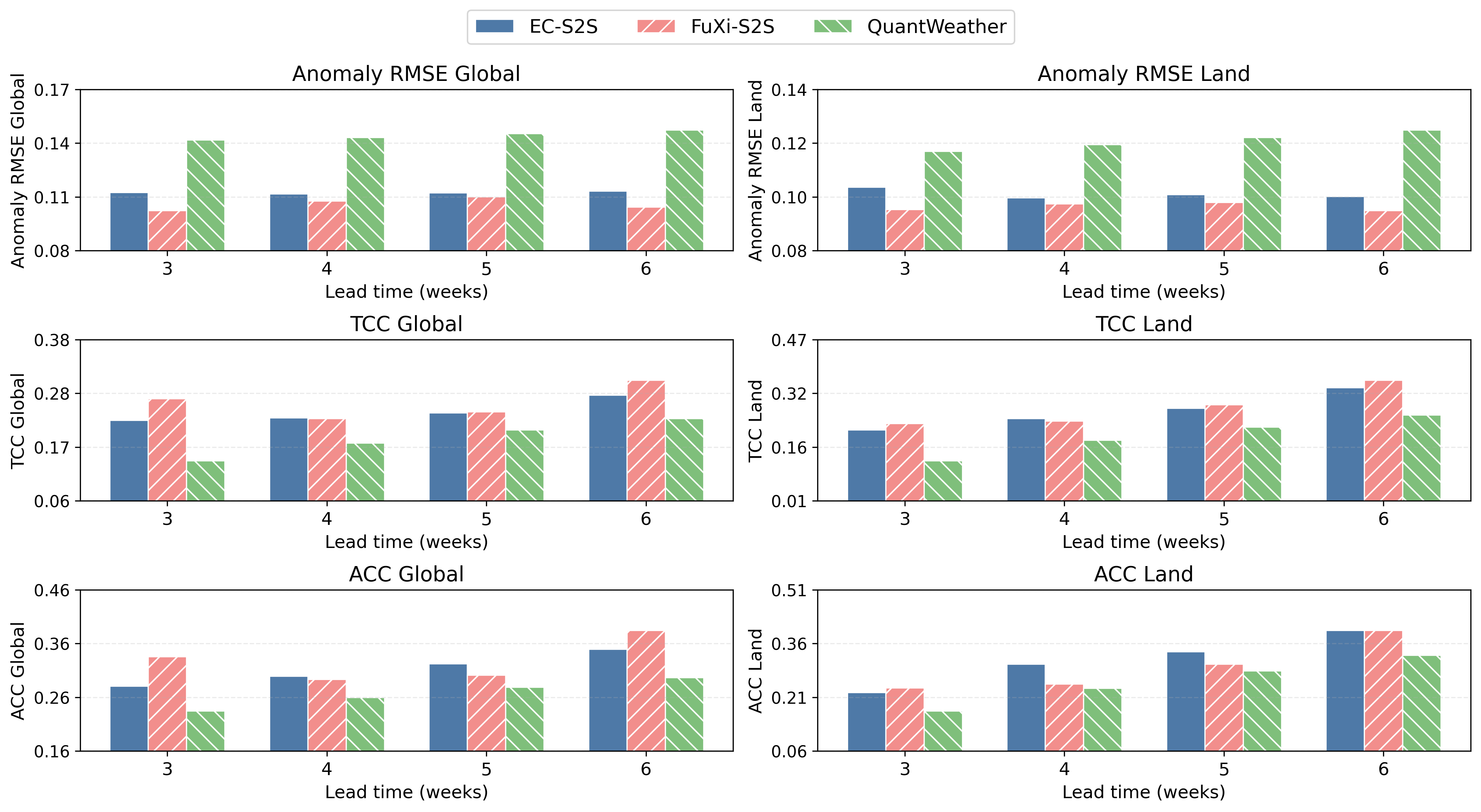}
        \caption{Member = 8}
        \label{app_fig:reg_m8}
    \end{subfigure}
    \caption{\small
    Comparison of anomaly-field RMSE, TCC, and ACC for regression forecasts using different numbers of inference members. The results are reported from Week 3 to Week 6 over global and land regions. Lower RMSE is better, while higher TCC and ACC indicate better performance.
    }
    \label{app_fig:reg_m1_m8}
\end{figure}

% \clearpage
% \input{checklist.tex}

\end{document}